\documentclass[10pt]{article}

\usepackage{amsmath}
\usepackage{amssymb}
\usepackage{graphicx}
\usepackage{cite}
\usepackage{color}

\usepackage[labelfont=bf,labelsep=period,justification=raggedright]{caption}

\makeatletter
\renewcommand{\@biblabel}[1]{\quad#1.}
\makeatother

\date{}

\begin{document}

\begin{flushleft}
{\Large
\textbf{Faster is More Different: Mean-Field Dynamics of Innovation Diffusion}
}
\\
Seung Ki Baek$^{1,\ast}$,
Xavier Durang$^{1}$,
Mina Kim$^{2}$
\\
\bf{1} School of Physics, Korea Institute for Advanced Study, Seoul 130-722,
Korea
\\
\bf{2} Department of Physics, University of Seoul, Seoul 130-743, Korea
\\
$\ast$ E-mail: seungki@kias.re.kr
\end{flushleft}

\section*{Abstract}
Based on a recent model of paradigm shifts by Bornholdt et al., we studied
mean-field opinion dynamics in an infinite population where an infinite
number of ideas compete simultaneously with their values publicly known. We
found that a highly innovative society is not characterized by
heavy concentration in highly valued ideas:
Rather, ideas are more broadly distributed in a
more innovative society with faster progress, provided that the rate of
adoption is constant, which suggests a positive correlation between
innovation and technological disparity. Furthermore, the distribution is
generally skewed in such a way that the fraction of innovators is
substantially smaller than has been believed in conventional
innovation-diffusion theory based on normality. Thus, the typical adoption
pattern is predicted to be asymmetric with slow saturation in the ideal
situation, which is compared with empirical data sets.

\section*{Introduction}

Pursuing new ideas is a fundamental characteristic of our modern society, where
brand-new goods are always ready to push their predecessors off the market.
Innovation is one of the most
important keywords to understand our society in this sense, as earlier
societies were shaped by traditional ideas to be
conserved in an unaltered form as much as possible. For this reason,
there have been extensive empirical economic and business studies
on how innovations get started, diffused and approved in a
society, and it is becoming an attractive topic in statistical physics as
well~\cite{perc,martins,guard,kim,souza,chae,born,holyst}.
In a classical work~\cite{rogers} about diffusion of innovations,
Rogers claimed
that there is a common pattern in innovation dynamics, that
people adopting an innovation are normally distributed in time. As a
result, the cumulative number of adopters is expected to show an $S$-shaped
pattern over time, which is described by the error function:
It grows slowly at first, expands rapidly at some point,
and then slowly saturates to 100\%. Deviation from the mean adoption time,
$\bar{t}$, over the entire population defines five adopter
categories such as innovators ($t<\bar{t}-2\sigma$, 2.5\%), early adopters
($\bar{t}-2\sigma<t<\bar{t}-\sigma$, 13.5\%), the early majority
($\bar{t}-\sigma<t<\bar{t}$, 34\%), the late majority
($\bar{t}<t<\bar{t}+\sigma$, 34\%), and laggards ($t>\bar{t}+\sigma$, 16\%),
where $\sigma$ is the standard deviation of adoption time.
If normality was true, it might reflect variations in
individual innovativeness, which is possibly an aggregate of numerous
random events and is normally distributed over the population.
However, this is a purely static picture of a non-communicating
population and it is an implausible description of an innovative
society.

At the same time, Rogers suggested a dynamic origin of this $S$-shaped
pattern by comparing it to an epidemic process. A relevant description is
then more likely to be a logistic function (see, e.g.,
Refs.~\cite{murray,mahajan,barnett}) than the error function. A logistic
function is
basically written as $h(t) = \frac{1}{2} \left( 1 + \tanh \frac{t}{2}
\right)$, which grows from zero to one as time $t$ goes from $-\infty$ to
$+\infty$. Here, the assumption is that there is a \emph{single}
innovation like a disease, diffusing into a passive population. However,
the problem with this approach is that ideas are evolving during the course of
adoption, and innovation researchers are already well aware that
people actively modify an adopted idea whenever it is possible and
necessary, which is termed re-invention~\cite{karn}
As a consequence, it is the rule rather than the exception
that every modified innovation may well compete with all its predecessors,
so the picture becomes more colorful than the dichotomy of a new idea versus
an old one. In short, this epidemic description does not capture the genuine
dynamic feature of innovations, and even more refined mathematical
approaches such as the Bass model do not overcome such
limitations~\cite{bass,mahajan,barnett}. This issue is also
deeply related to the pro-innovation bias of diffusion
research~\cite{rogers}, which means that one tends to overlook such an
innovation that dies out by rejection or replaced by a better one.
Although there have been statistical-physical approaches to introduce
many competing ideas into the dynamics of
innovation~\cite{guard,souza,kim,chae}, they are rather focused on scaling
behavior under specific stochastic rules than comparing the findings
with empirical observations.

To sum up, analytic concepts are lacking to
explain actual patterns of innovation diffusion as a fully dynamic
process with a multitude of ideas competing simultaneously. For this
reason, we consider simple ideal competition among ideas whose
values are so well-defined that everyone can adopt a better idea as soon as
she encounters it,
without any barriers against the diffusion of innovations. Even if this
picture is unrealistic, it is theoretically intriguing, and
can serve as a reference point to start with when assessing
innovations in practice.
In particular, our results suggest that the interplay of
adoption and exploration must be considered to achieve a plausible minimalist
description, which leads to neither normal nor logistic but slightly
skewed behavior as a signature of an ideal innovative society. This simple
explanation is in contrast to many variants of the logistic
growth model that describe asymmetry in empirical $S$-shaped
patterns~\cite{mahajan,barnett}.
Moreover, the analysis tells us that the speed of progress in ideas is
coupled to how broadly ideas are distributed in the society: a fast
innovating society tends to be accompanied by a broad spectrum of ideas, some
of which can be far from state-of-the-art.
It should be kept in mind
that the term `ideal' is absolutely unrelated to any judgments of value
concerning the phenomena that we are investigating but only means that
we are considering a conceptual construct that can be pursued analytically.

\section*{Methods of Analysis}

Following Ref.~\cite{born}, we assume that every idea is assigned a scalar
value $x$ representing its quality. This automatically implies that
this quantity is transitive without any cyclic dominance among ideas, and
the strict dominance relationship between any pair of distinct ideas prevents
people from revisiting old ideas. A difference from
Ref.~\cite{born} is that $x$ can take any \emph{real} value, not only an
integer.
Let $P(x,t) dx$ denote the fraction of the population choosing ideas between
$x$ and $x+dx$ at time $t$. We then call $P(x,t)$ a probability density
function (pdf) of idea $x$.
Our population dynamics approach on the mean-field level suggests that
the relative growth rate $\frac{1}{P(x,t)} \frac{\partial P(x,t)}{\partial
t}$ is proportional to the fraction of those with $x'<x$ as they are
potential adopters of $x$.
This fraction is, by definition, the cumulative distribution function (cdf)
$C(x,t) \equiv \int_{-\infty}^x P(x';t) dx'$ and we thus have
\begin{equation}
\frac{\partial P(x,t)}{\partial t} = k [C(x,t) - \bar{C}(t)] P(x,t),
\label{eq:dp}
\end{equation}
where $k$ is a positive proportionality constant representing the rate of
adoption, which can be set as unity by using a rescaled time $\tilde{t} =
kt$, and $\bar{C}(t)$ is the average of $C(x,t)$ over the population.
Note that the total probability is always conserved
because $\int \frac{\partial
P(x,t)}{\partial t} dx = k [\bar{C}(t) -\bar{C}(t)] = 0$~\cite{rd}.
An alternative way to derive Eq.~(\ref{eq:dp}) is to start from a master
equation~\cite{noh}:
\[ \frac{\partial P(x,t)}{\partial t}= \frac{k}{2} \int_{-\infty}^{x}{dy
P(y,t)P(x,t)}- \frac{k}{2}\int_{x}^{\infty}{dy P(y,t)P(x,t)}, \]
where the first term describes an inflow adopting $x$ and the second term
describes an outflow adopting higher values than $x$. It could also be
modified by inserting suitable kernel functions into the integrals.

\begin{figure}[!ht]
\includegraphics[width=0.45\textwidth]{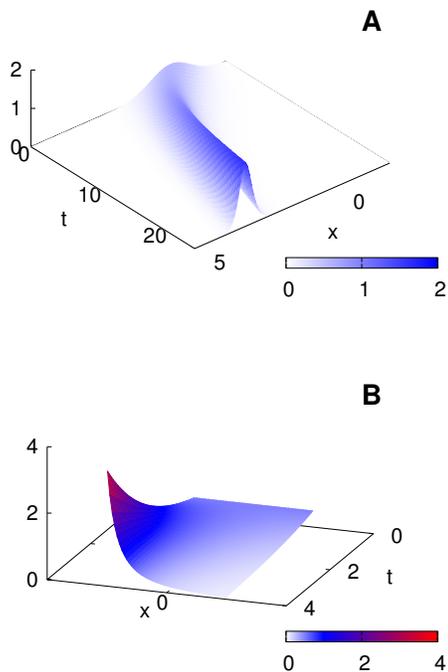}
\caption{Adoption-only dynamics with different initial
conditions. (A) The normal distribution with unit variance
[Eq.~(\ref{eq:gauss})]. (B) The box distribution defined on $-1<x<1$
[Eq.~(\ref{eq:box})].}
\label{fig:gauss}
\end{figure}

An integration by parts yields
\[ \bar{C} (t) \equiv 
\int_{-\infty}^\infty C(x,t) \frac{\partial C(x,t)}{\partial x} dx =
\left[ \frac{1}{2} C^2(x,t) \right]_{x=-\infty}^{\infty} = \frac{1}{2} \]
since $C(x=-\infty;t) = 0$ and $C(x=\infty;t) = 1$.
It is convenient to rewrite Eq.~(\ref{eq:dp}) only in terms of $C(x,t)$:
\begin{equation}
\frac{\partial^2 C(x,t)}{\partial t~ \partial x} = k\left[C(x,t) - \frac{1}{2}
\right] \frac{\partial C(x,t)}{\partial x}.
\label{eq:dp2}
\end{equation}
A stationary state with $\frac{\partial C(x,t)}{\partial t}=0$ requires
$\frac{\partial C(x,t)}{\partial x} = 0$ in
Eq.~(\ref{eq:dp2}) since $C(x,t) \neq \frac{1}{2}$ in general
due to the boundary condition at $x=\pm \infty$.
The vanishing derivative with respect to $x$ means that $P(x,t) =
\delta (x-x_1)$ with some constant $x_1$, which should be the highest value
in the initial pdf with a compact support
such that $P(x,t_0) > 0$ only for $x_0<x<x_1$ at the initial time $t_0$.
To proceed to the general solution, let us rewrite
Eq.~(\ref{eq:dp2}) as
\begin{equation}
\frac{\partial}{\partial x} \left[ \frac{\partial
\hat{C}(x,\tilde{t})}{\partial \tilde{t}} -
\frac{1}{2} \hat{C}^2(x,\tilde{t}) \right] = 0,
\label{eq:dp3}
\end{equation}
where $\hat{C}(x,\tilde{t}) \equiv C(x,\tilde{t}) - \frac{1}{2}$ with the
rescaled time $\tilde{t} = kt$.
Clearly, Eq.~(\ref{eq:dp3}) implies that
the expression inside the brackets is a function of $t$ and independent
of $x$. Inserting the
boundary condition at $x=\pm \infty$, the expression inside the bracket
is $-1/8$ at every $\tilde{t}$. This means that
the equation to be solved is the following:
\begin{equation}
\frac{\partial \hat{C}(x,\tilde{t})}{\partial \tilde{t}} - \frac{1}{2}
\hat{C}^2(x,\tilde{t}) = -\frac{1}{8}.
\label{eq:eighth}
\end{equation}
The solution can be found as
\begin{equation}
\hat{C}(x,\tilde{t}) = \frac{1}{2} \tanh \left[ g(x) - \tilde{t}/4
\right]
\label{eq:cdf}
\end{equation}
with a certain function $g(x)$.
The definition of $\hat{C}(x,t)$ requires $d g(x)/ d x \ge 0$ with
$g(x \rightarrow +\infty) = +\infty$ and
$g(x \rightarrow -\infty) = -\infty$.
In terms of the pdf, it means that
\begin{equation}
P(x,t) = \frac{\partial \hat{C}(x,t)}{\partial x} = \frac{1}{2}
\left[ \frac{d g(x)}{d x} \right] {\rm
sech}^2 \left[ g(x) - \frac{kt}{4} \right],
\label{eq:sol}
\end{equation}
where $\hat{C}(x,t) \equiv C(x,t) - \frac{1}{2}$ and
$g(x)$ is an arbitrary function satisfying
$d g(x)/ d x \ge 0$ with
$g(x \rightarrow +\infty) = +\infty$ and
$g(x \rightarrow -\infty) = -\infty$.
It can be readily checked that it contains the
stationary delta function as a special case.
If the initial distribution at $t=0$ is a normal distribution with unit
variance,
\[ C(x,0) = \frac{1}{2} \left[ 1 + {\rm erf}(x) \right], \]
and the time evolution is determined as
\begin{equation}
P(x,t) =
\frac{e^{-x^2}}{\sqrt{\pi}[1-{\rm erf}^2(x)]}{\rm sech}^2\left\{ g(x)
 - \frac{t}{4}\right\},
\label{eq:gauss}
\end{equation}
where ${\rm erf}(x)$ is the error function and $g(x) =
{\rm arctanh}\left[ {\rm erf}\left(x\right)\right]$. The speed of this wave is
$v(x) = x/[4 g(x)] = g^{-1}(t/4)/t$,
which decreases over time. As the speed decreases, the wave
becomes sharper [Fig.~\ref{fig:gauss}(a)].
As another example, we take a box distribution defined on the interval
between $x=-1$ and $+1$ as our initial pdf $P(x,0)$. Then we have
\[g(x) = {\rm atanh}\left\{ x[H(x+1)-H(x-1)] - H(-x-1) + H(x-1)] \right\},\]
where $H(x)$ is the Heaviside step function. The solution is given as
\begin{equation}
P(x,t) = \left\{
\begin{array}{ll}
\frac{{\rm sech}^2\left[ {\rm arctanh}(x) - \frac{t}{4} \right]}{2-2x^2} &
\mbox{~~if~~}-1<x<1,\\
0 & \mbox{~~otherwise.}
\end{array}
\right.
\label{eq:box}
\end{equation}
As time goes by, it converges to a delta peak at $x=1$
[Fig.~\ref{fig:gauss}(b)].

Let us return to the general solution [Eq.~(\ref{eq:sol})].
For any $g(x)$ and $x_0$, the fraction of the
population having passed this innovation level $x_0$, i.e., $1-C(x_0,t)$,
increases as a logistic function of $t$.
However, it should be noted that our starting point was not meant to be the
logistic growth model.
The time evolution
of $P(x,t)$ is fully determined once $g(x)$ is given by the initial
condition, suggesting that innovation history is
already determined at the starting point
as long as the rate of adoption $k$ remains unaltered. If the
initial condition is nonzero only over a finite range of $x \in [x_1, x_2]$,
for example, $P(x,t)$ always evolves to a delta function at $x_2$.
This deficiency makes it difficult to gain insight on the innovation
dynamics from the current formulation, revealing its incompleteness.

The reason is that our current formulation does not include any generative
mechanism for innovations.
Therefore, we add another term to the adoption dynamics considered so
far. It could be argued that individual exploration for different ideas can be
modeled more or less by a Brownian random walk along the $x$-axis:
\begin{equation}
\frac{\partial P(x,t)}{\partial t} = D \frac{\partial^2 P(x,t)}{\partial
x^2},
\label{eq:brown}
\end{equation}
where $D$ is a measure of exploratory efforts.
Because it yields a normal distribution with variance $2Dt$, this
could be interpreted as invoking the classical idea of normality in the
diffusion of innovations, but this normality enters as a consequence of the
dynamic exploration process rather than a static trait. It also expresses a
conservative viewpoint that an individual alone achieves only small
modifications that may even degenerate equally. This is obviously a
huge simplification about the human mind, but we shall be content with such a
minimalist description at the moment.
Adding this exploratory mechanism to the adoption,
the resulting equation is written as
\begin{equation}
\frac{\partial^2 \hat{C}(x,t)}{\partial t \partial x} = k \hat{C}(x,t)
\frac{\partial \hat{C}(x,t)}{\partial x} + D \frac{\partial^3
\hat{C}(x,t)}{\partial x^3}.
\label{eq:diff}
\end{equation}
By rescaling $\tilde{t} = kt$ and $\tilde{x} = x \sqrt{k/D}$,
we set both parameters $k$ and $D$ as unity.
Notably, Eq.~(\ref{eq:diff}) does not have a stationary
solution for the following reason: When $\partial \hat{C}(x,t) / \partial t
= 0$, the solution for Eq.~(\ref{eq:diff}) is given as Weierstrass' elliptic
function, which is even and does not satisfy the
boundary condition of $\hat{C}(x,t)$ at $x=\pm \infty$.
This might look counter-intuitive at first glance as the pdf
tends to converge to a single point due to
adoption, which could be balanced by exploration. However, a more
correct picture is that the pdf converges to a \emph{higher} position
than the center, so it gradually moves upward via exploration instead of
staying at a fixed position. This notion turns out to be plausible
as will be explained shortly below.

If we consider the boundary condition,
the actual equation to solve here is given as
\begin{equation}
\frac{\partial \hat{C}(\tilde{x},\tilde{t})}{\partial \tilde{t}}
= \frac{1}{2} \hat{C}^2(\tilde{x},\tilde{t}) -\frac{1}{8}
+ \frac{\partial^2 \hat{C}(\tilde{x},\tilde{t})}{\partial \tilde{x}^2},
\label{eq:fisher}
\end{equation}
which can be shown identical to Fisher's equation~\cite{fisher} by simply
changing the variables. Fisher's equation was originally devised to describe
the frequency of a single mutant gene in a one-dimensional population rather
than a cdf, and it is interesting that the same equation arises in
the context of an infinite series of mutants in an infinite-dimensional
(i.e., mean-field) population. This equation has been extensively studied in
biology and physics as one of the simplest reaction-diffusion
systems~\cite{murray,saarloos,fort}. We only mention the basics of the known
results about Fisher's equation and those who are interested in
comprehensive discussions may refer to Ref.~\cite{murray} and references
therein.
 
Equation~(\ref{eq:fisher}) admits traveling wave solutions,
and preserves the shapes during propagation. The traveling wave solutions are
stable against small perturbations within a finite domain, moving with the
waves.
Each speed builds up a unique wave shape, and speed $v$ is
determined by the tail of the initial cdf in the following manner: If
$C(\tilde{x},\tilde{t}_0) \sim e^{-a\tilde{x}}$ with $a>0$ as $\tilde{x}
\rightarrow \infty$ at initial time $\tilde{t}_0$, the speed of the
wavefront asymptotically converges to $v =
\sqrt{Dk/2}\left(a +a^{-1} \right)$ when $a\le 1$, and $v=v_{\rm
min}=\sqrt{2Dk}$ when $a>1$. In short, a longer tail leads to a faster
propagating wave. Even if an initial pdf has bounded support, i.e.,
$P(x,t_0)> 0$ only for $x<x_1$, a traveling wave solution will develop
with $v=v_{\rm min}$ instead of a delta function.
The information on the initial condition other than the tail exponent
becomes irrelevant in the
asymptotic limit due to the random-walk process. There is no
traveling wave solution below $v_{\rm min}$, which is consistent with the
impossibility of a stationary solution as stated above.
Another important feature is that the characteristic width $w$ of the
wavefront is proportional to $\sqrt{D/k}$ because $D$ and $k$ compete to
determine width. In contrast, speed is expressed as $v \propto
\sqrt{Dk}$ as both the mechanisms of exploration and adoption
make positive contributions. As a consequence, the characteristic time for a
wavefront to pass through a particular point $x$ is not sensitive to $D$
because $w/v \sim k^{-1}$.

\begin{figure}[!ht]
\includegraphics[width=0.45\textwidth]{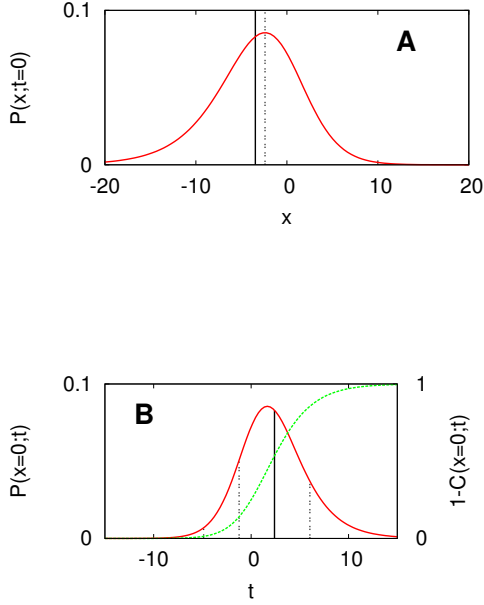}
\caption{Shape of a traveling wave [Eq.~(\ref{eq:special})] resulting from
Fisher's equation.
(A) $P(x,t) = \partial C(x,t)/ \partial x$ at
$t=t_0$ with $D=k=1$. The solid vertical line is the mean, and the
dotted vertical line is the mode of the pdf.
(B) Temporal pattern of adopting an innovation $x=0$ with $D=k=1$ and
$t_0=0$.
The solid (red) curve $P(x,t)$ shows how the fraction of the
population with $x=0$ changes over time,
whereas the dotted (green) curve $1-C(x,t)$ shows the
fraction that has adopted $x \ge 0$ as a function
of time. The solid vertical line
is the mean adoption time $\bar{t}$, and the dotted vertical lines represent
$\bar{t}-2\sigma$, $\bar{t}-\sigma$, and $\bar{t}+\sigma$, respectively, to
distinguish the five adopter categories.
}
\label{fig:diff}
\end{figure}

A fully analytic expression
for a specific velocity $v = \frac{5}{2\sqrt{3}}\sqrt{Dk} \approx 1.02~
v_{\rm min}$ is available as:
\begin{eqnarray}
C(x,t) &=& \frac{1}{4} \left\{ 3 + 2\tanh\left[
\frac{x}{4\sqrt{3D/k}} - \frac{5k}{24}(t-t_0) \right]\right.\nonumber\\
&&\left. -\tanh^2 \left[ \frac{x}{4\sqrt{3D/k}} -
\frac{5k}{24}(t-t_0)\right]
\right\},
\label{eq:special}
\end{eqnarray}
where $t_0$ is a reference point in time~\cite{ablo,kali}.
As this expression is handy to maintain qualitative features
unaltered, we will focus on this solution to observe differences from the
normal or logistic descriptions. The numbers presented here should be
taken as indicating qualitative features of the solution, and not as
universal values for arbitrary $v$.
The shape of the wave $P(x,t)$ is obtained by differentiating
Eq.~(\ref{eq:special}) with respect to $x$,
which is shown in Fig.~\ref{fig:diff}(a) at $t=t_0$.
As is clearly shown there,
this pdf is not symmetric but skewed
negatively, i.e., with a longer tail on the left side. The skewness is
quantified from the second and third moments as $\gamma_1
\approx -0.5772$. Due to this skewness, while the mean is $\bar{x} \approx
-3.464\sqrt{D/k}$, the maximum is located at $x^\ast \approx - 2.401
\sqrt{D/k}$. Consequently, the most commonly observed idea tends to lead us
to overestimate the population mean.
Recall the five categories defined with respect to the
\emph{mean} adoption time, which is given by our $P(x,t)$ as
$\bar{t} = \left[ \int_{-\infty}^\infty t P(x=0;t)
dt \right] \Big/ \left[ \int_{-\infty}^\infty P(x=0;t) dt \right] =
\frac{12}{5}$,
when $t_0=0$ and the idea to adopt has value $x=0$
[Fig.~\ref{fig:diff}(b)]. The standard deviation around $\bar{t}$ is $\sigma
\approx 3.632$, from which we can compute fractions of the five categories
as $1.36\%$ (innovators), $12.66\%$ (early adopters), $39.42\%$ (early
majority), $32.13\%$ (late majority), and $14.43\%$ (laggards). Note
that the fraction of innovators is only one half of the existing estimate
based on the normality.
This is due to the inherent skewness of the pdf
as a solution for this dynamics.
The shape of $P(x,t) \propto C(x+dx,t)-C(x,t)$ in
Fig.~\ref{fig:diff}(b) can also be interpreted as the typical fate of idea
$x$, spread by adoption but soon dominated by its descendant $x+dx$.

\section*{Empirical Results}

Although Eq.~(\ref{eq:special}) describes only a special case of a specific
velocity, we can verify whether it fits to the empirical data set found in
Ref.~\cite{rogers}.
Recall that a traveling wave with $v=v_{\rm min}$ emerges from any initial
pdf with a sufficiently short tail, which we presume is close to reality in
many cases.
Therefore, it would be useful to directly work with this solution,
but it is more difficult to handle than the analytic solution
Eq.~(\ref{eq:special}) for practical purposes. Fortunately, the analytic
solution shows little difference in its shape compared to the
solution with $v_{\rm min}$. Thus, we work with Eq.~(\ref{eq:special})
to interpret two different data sets : 
the cumulative number of publications of innovation and
the broadband penetration rates in European countries.

\subsection*{Publications of innovation}
\begin{figure}[!ht]
\includegraphics[width=0.45\textwidth]{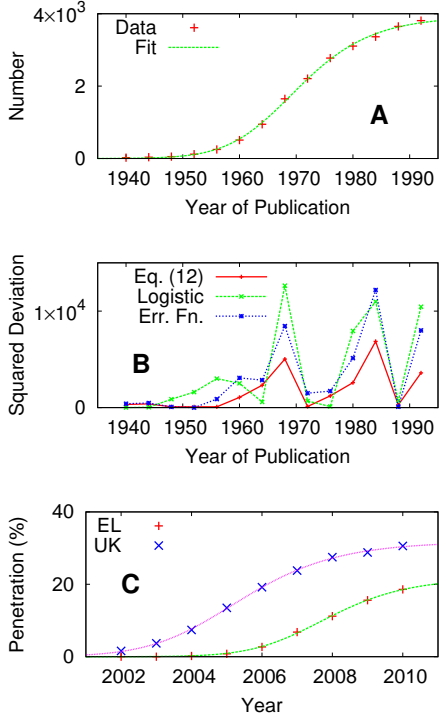}
\caption{Comparison of Eq.~(\ref{eq:special}) with empirical data. (A) 
Cumulative number of publications on the diffusion of
innovations, excerpted from Ref.~\cite{rogers}. The curve is obtained by
fitting a functional form $N(t) = N_s[1-C(x,t)]$
[see Eq.~(\ref{eq:special})] to the data points where $N_s$ is the
saturation number at $t
\rightarrow \infty$. The fitting parameters are $(N_s,t_0,k) =
(3294,1964,0.32)$. (B) The same data shows larger deviations when
fitted with the logistic function $N_s \left\{ 1+\tanh[k(t-t_0)]
\right\}/2$ (green) or the error function $N_s \left\{ 1+ {\rm erf}[k(t-t_0)]
\right\}/2$ (blue). Their best fitting parameters are
$(N_s,t_0,k)=(3797,1970,0.086)$ and $(3769,1970,0.072)$, respectively.
(C) Broadband penetration rates in Greece and the United
Kingdom (UK) from Eurostat~\cite{post}. The curves were obtained in the same
way as above with Eq.~(\ref{eq:special}), yielding $(N_s,t_0,k) =
(21.7,2007,1.84)$ for Greece and
$(31.7,2004,1.53)$ for the UK.}
\label{fig:pub}
\end{figure}

The data set in Fig.~\ref{fig:pub}(a) shows the cumulative numbers of
publications on the diffusion of
innovations every 4 years from 1940 to 1996.
As we approach the late 1990s, the rate of increase decreases, but it is not
symmetric with the early take-off around the 1960s. That
is, the shape is slightly skewed as our theory suggests
[Fig.~\ref{fig:diff}(b)]. The curve in Fig.~\ref{fig:pub}(a) shows our fit of
Eq.~(\ref{eq:special}) to the data set by the least-squares method.
Although the attempt is quite cavalier, the agreement with the data points
is excellent. When compared to fittings with the error function and the
logistic function, this functional form actually provides a better
explanation, in the sense that the sum of squared deviations becomes one
half of each of theirs [Fig.~\ref{fig:pub}(b)].
From this fitting, we can estimate the rate of
adoption $k \approx 0.32$. Plugging this value into Eq.~(\ref{eq:special}),
we suggest that the relevant time scale of adopting the diffusion concept of
innovations amounts to $24/(5k) \approx 15$ years.
One could argue from this excellent fit that the research field is close to the
ideal situation that we have considered: researchers are relatively
open-minded about new ideas and their communication is not much restricted
by geographic factors. Based on this idea, the
deviation of empirical adoption patterns from the predicted curve can
serve as an indicator to quantify barriers against diffusion of innovations.
For example, a classical study of diffusion research on the hybrid corn in
Iowa~\cite{rogers} shows a positively skewed pdf contrary to the prediction,
which may hint at the strong resistance by the farmers to the new idea at
the early stage.

\subsection*{Broadband penetration in Europe}
Our second example in Fig.~\ref{fig:pub}(c) shows broadband penetration
rates in European countries, as published by Eurostat~\cite{post}. This
quantity
means the number of high-speed connections ($\ge$ 144 Kbits/s) per 100
inhabitants. The figure tells us that the broadband penetration in Greece
started about 3 years later than that in the UK, and its
saturation level in the future will be $10\%$ lower than that of the UK.
Despite these differences, the relevant time scales of adoption are
estimated to be about 3 years for both countries.

\begin{figure}[!ht]
\includegraphics[width=0.45\textwidth]{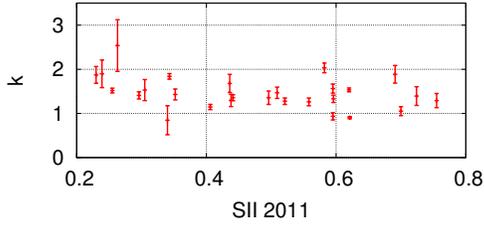}
\caption{Summary innovation index (SII) versus the rates of
adoption in the European Union (EU) member countries.
}
\label{fig:sii}
\end{figure}

In fact, the rates of adoption, evaluated from the broadband penetration
rates, do not change much across European countries.
Table~\ref{table:fit} shows the least-square fitting results of
Eq.~(\ref{eq:special}) to the broadband penetration rates from 2002 to 2010
in EU member countries~\cite{post}. Note that the values in column $t_0$
are relative to 2002.
In Fig.~\ref{fig:sii}, we plot the resulting $k$ values in
Table~\ref{table:fit}.
The horizontal axis represents the summary innovation
index (SII), which has been developed to assess aggregate national
innovation performance of the EU member countries~\cite{ius}.
It is a composite index showing how many relevant indicators such as
education, employment and R\&D are above or below EU averages.
Figure~\ref{fig:sii} suggests that the differences in innovativeness
measured by the SII cannot be explained by the differences in the
rates of adoption. Therefore, if we use the SII as a proxy variable for
measuring speed $v \propto \sqrt{Dk}$, the differences in the SII should be
explained by variations in the measure $D$ of exploration activity.

If $k$ is uniform, our model predicts that
more diverse values of $x$ will be observed in a society where innovation
occurs faster because both $v$ and $w$ scale as $\sqrt{D}$. The abundance of
laggards with low $x$ results from the fast innovation but also fuels it
as market potential, and both effects are incorporated in the
solution.
 
\section*{Discussion and Summary}

In summary, we have studied an ideal innovative society where a better
idea has a better chance to diffuse into the population. Our model is
characterized by competition among an infinite number of ideas.
In the presence of an adoption mechanism only,
we are able to find the full solution exhibiting logistic
behavior, but it is a purely deterministic view leaving the concept of
innovation obscure.
By adding another term for exploratory behavior, which connects to the classical
idea of normality, we have found traveling wave solutions as described by
Fisher's equation, whose velocity is proportional to the square root of
exploration activity $D$ times the rate of adoption $k$. At the same time,
its width is proportional to $\sqrt{D/k}$ due to the competition of
adoption and exploration.
Incorporating both the normal and logistic features,
the shape of the solution is neither normal
nor logistic but negatively skewed, leading to a discrepancy between the
mean and the mode as well as a significantly smaller size estimate of
innovators compared to that of the conventional theory. It is compared with the
asymmetry in empirical adoption patterns and proposed as a reference
point to assess the effectiveness in diffusion of innovations.
Furthermore, as the rates of adoption do not vary much across countries,
we predict a tendency for the width of a distribution to be positively
correlated with the overall speed of innovations.

\section*{Acknowledgments}
S.K.B. is grateful to Beom Jun Kim and Jae-Suk Yang for stimulating
conversations.
We thank KIAS Center for Advanced Computation for providing computing
resources.


\section*{Figure Legends}

\section*{Tables}
\begin{table}
\caption{Fitting results of Eq.~(\ref{eq:special}) to the
broadband penetration rates from 2002 to 2010 in EU member
countries.}
\begin{tabular}{cccccccc}
Country & $N_s$ & $t_0$ & $k$ & Country & $N_s$ & $t_0$ & $k$ \\
BE & 33.5 $\pm$ 0.4 & 0.49 $\pm$ 0.04 & 0.90 $\pm$ 0.02 & LU & 34.8 $\pm$ 0.9 & 2.58 $\pm$ 0.08 & 1.56 $\pm$ 0.10 \\
BG & 15.5 $\pm$ 1.0 & 4.43 $\pm$ 0.11 & 1.90 $\pm$ 0.31 & HU & 22.0 $\pm$ 1.0 & 3.36 $\pm$ 0.10 & 1.43 $\pm$ 0.12 \\
CZ & 21.4 $\pm$ 1.1 & 3.37 $\pm$ 0.13 & 1.68 $\pm$ 0.21 & MT & 43.4 $\pm$16.9 & 3.74 $\pm$ 1.15 & 0.85 $\pm$ 0.33 \\
DK & 40.7 $\pm$ 2.0 & 0.99 $\pm$ 0.21 & 1.39 $\pm$ 0.21 & NL & 40.5 $\pm$ 0.9 & 1.06 $\pm$ 0.08 & 1.33 $\pm$ 0.08 \\
DE & 38.5 $\pm$ 2.5 & 2.67 $\pm$ 0.18 & 1.05 $\pm$ 0.10 & AT & 26.9 $\pm$ 1.2 & 1.11 $\pm$ 0.14 & 0.93 $\pm$ 0.08 \\
EE & 28.5 $\pm$ 1.2 & 1.99 $\pm$ 0.13 & 1.36 $\pm$ 0.15 & PL & 18.5 $\pm$ 0.8 & 4.31 $\pm$ 0.08 & 1.41 $\pm$ 0.08 \\
IE & 23.7 $\pm$ 0.5 & 3.38 $\pm$ 0.05 & 2.03 $\pm$ 0.11 & PT & 19.8 $\pm$ 0.9 & 1.48 $\pm$ 0.15 & 1.30 $\pm$ 0.14 \\
EL & 21.7 $\pm$ 0.5 & 4.75 $\pm$ 0.04 & 1.84 $\pm$ 0.06 & RO & 14.0 $\pm$ 0.7 & 4.25 $\pm$ 0.19 & 2.54 $\pm$ 0.58 \\
ES & 24.9 $\pm$ 0.6 & 1.84 $\pm$ 0.08 & 1.15 $\pm$ 0.06 & SI & 26.8 $\pm$ 0.8 & 2.82 $\pm$ 0.07 & 1.28 $\pm$ 0.07 \\
FR & 33.9 $\pm$ 1.1 & 2.10 $\pm$ 0.10 & 1.26 $\pm$ 0.09 & SK & 19.1 $\pm$ 2.0 & 4.39 $\pm$ 0.20 & 1.53 $\pm$ 0.24 \\
IT & 22.3 $\pm$ 0.5 & 1.93 $\pm$ 0.07 & 1.36 $\pm$ 0.07 & FI & 31.1 $\pm$ 0.9 & 1.36 $\pm$ 0.13 & 1.89 $\pm$ 0.20 \\
CY & 28.0 $\pm$ 1.6 & 4.12 $\pm$ 0.11 & 1.47 $\pm$ 0.13 & SE & 35.4 $\pm$ 1.7 & 1.26 $\pm$ 0.17 & 1.29 $\pm$ 0.16 \\
LV & 20.1 $\pm$ 0.8 & 3.47 $\pm$ 0.10 & 1.87 $\pm$ 0.19 & UK & 31.7 $\pm$ 0.3 & 2.05 $\pm$ 0.04 & 1.53 $\pm$ 0.04 \\
LT & 21.5 $\pm$ 0.4 & 3.12 $\pm$ 0.04 & 1.52 $\pm$ 0.05 & & & & \\
\end{tabular}
\label{table:fit}
\end{table}

\end{document}